\documentclass[twocolumn,showpacs,preprintnumbers,prb]{revtex4}

\usepackage{amssymb}
\usepackage[dvips]{color}
\usepackage{epsfig}
\usepackage{dcolumn}
\usepackage{bm}

\def\G{\Gamma}

\def\e{\epsilon}

\def\s{\sigma}
\def\S{\Sigma}
\def\wid{\widetilde}

\begin{document}

\title{Quantum Dot as a Spin--Current Diode: a spin-dependent transport study}
\author{F. M. Souza,$^{1,2}$ J. C. Egues,$^{3}$ and A. P. Jauho$^{4}$}
\affiliation{$^1$ Instituto de Estudos Superiores da Amaz{\^o}nia,
66055-260, Bel{\'e}m, Par{\'a}, Brazil \\ $^2$ Grupo de F{\'i}sica
de Materiais da Amaz{\^o}nia, Departamento de F{\'i}sica,
Universidade Federal do Par{\'a}, 66075-110, Bel{\'e}m, Par{\'a}, Brazil \\
$^3$ Departamento de F\'{\i}sica e Inform\'{a}tica, Instituto de
F\'{\i}sica de S\~{a}o Carlos, Universidade de S\~{a}o Paulo,
13560-970, S\~{a}o Carlos, S\~{a}o Paulo, Brazil \\ $^4$ MIC -
Department of Micro and Nanotechnology, NanoDTU, Technical
University of Denmark, {\O}rsteds Plads, Bldg. 345E, DK-2800 Kgs.
Lyngby, Denmark } \keywords{spintronics, spin accumulation, Coulomb
blockade} \pacs{PACS number}

\begin{abstract}
We report a study of spin dependent transport in a system composed
of a quantum dot coupled to a normal metal lead and a ferromagnetic
lead (NM-QD-FM). We use the master equation approach to calculate
the spin-resolved currents in the presence of an external bias and
an intra-dot Coulomb interaction. We find that for a range of
positive external biases (current flow from the normal metal to the
ferromagnet) the current polarization
$\wp=(I_\uparrow-I_\downarrow)/(I_\uparrow+I_\downarrow)$ is
suppressed to zero, while for the corresponding negative biases
(current flow from the ferromagnet to the normal metal) $\wp$
attains a relative maximum value. The system thus operates as a
rectifier for spin--current polarization. This effect follows from
an interplay between Coulomb interaction and nonequilibrium spin
accumulation in the dot. In the parameter range considered, we also
show that the above results can be obtained via nonequilibrium Green
functions within a Hartree-Fock type approximation.
\end{abstract}

\volumeyear{year} \volumenumber{number} \issuenumber{number}
\eid{identifier}
\date[Date: ]{\today}
\maketitle

\section{Introduction}

Polarized transport in spin-dependent nanostructures is a subject of
intense study in the emerging field of spintronics,\cite{overview}
due to its relevance to the development of novel spin-based
devices.\cite{dl98,hae01,mw05} In addition, transport through QDs
provides information about fundamental physical phenomena in
spin-dependent and strongly correlated systems, such as the Kondo
effect,\cite{rs06,jm05,yu05,jm03,pz02} the Coulomb- and
spin-blockade effects,\cite{fe06,iw05,ac04,jb98,st98} spin valve
effect and tunnelling magnetoresistance
(TMR),\cite{kw06,iw06_2,jv05,iw05_2,fms04,fms02,rl03,iw03,wr03,wr01,jk03,mb04}
etc. Novel spin filters and pumps have also been proposed using QDs
coupled to normal metal leads.\cite{patrik,hanres,ec05} A system of
particular interest in this context comprises a quantum dot or a
metallic grain coupled to ferromagnetic leads. The ferromagnetism of
the leads introduces spin-dependent tunneling rates between the
leads and the central region. This results in a nonzero net spin in
the central region for asymmetric magnetization geometries. This
effect is called spin accumulation or spin
imbalance.\cite{ab99,hi99,jm02} It has been shown that spin
accumulation affects several transport properties, such as
magnetoresistance,\cite{iw05_2,iw06} (negative) differential
resistance\cite{fe06,iw06} and the zero-bias
anomaly.\cite{jm03,iw05,iw05_2} In addition it provides a way to
generate and control the current spin polarization via gates or bias
voltages,\cite{jw05,wk02} which is one of the main goals within
spintronics.

Systems composed of a nonmagnetic lead and a ferromagnetic lead with
a quantum dot or a quantum wire as spacer have been analyzed
recently. It was pointed out that if the spacer is a dot and the
ferromagnetic lead is half-metallic, a novel mesoscopic
current-diode effect arises.\cite{mw05,iw06,rs04} Spin-current
rectification was also predicted in an asymmetric system composed of
a ferromagnetic (Fe or Ni) and nonmagnetic (Au or Pd) contacts
coupled to each other via a molecular wire.\cite{hd06} Additionally,
it was pointed out that a NM-QD-FM system can operate as a
spin-filter and as a spin-diode.\cite{aas05} In Ref.
[\onlinecite{aas05}] the authors use the bias voltage to change the
resonance position of the dot level with respect to the spin-split
density of states of the ferromagnetic lead. This gives rise to
spin-dependent currents.

Here we study spin-resolved currents in a single-level quantum dot
attached to a nonmagnetic lead (``left lead") and to a ferromagnetic
lead (``right lead"), Fig. 1. As we shall show, the magnetic
asymmetry between the left and right terminals results in a
rectification of the current polarization for a particular bias
range for which the single electron channel $\e_d$ is on-resonance,
and the double-occupancy channel $\e_d+U$ is off-resonance. More
precisely, when the nonmagnetic lead operates as an emitter and the
ferromagnetic lead as a collector, defined as the positive bias
($eV>0$), the current is unpolarized. In contrast, when the
ferromagnetic lead is the emitter and the nonmagnetic lead is the
collector (negative bias) a spin-polarized current arises.
Importantly, this rectification occurs only in this particular bias
range, as we shall demonstrate both analytically and numerically.
This is attributed to an interplay between nonequilibrium spin
accumulation and Coulomb interaction within the dot. For high enough
bias voltages, the current polarization is essentially symmetric
with respect to the bias, and no rectification is found.

In the main body of the text we employ the master-equation approach
of Glazman and Matveev\cite{glaz-mat} to describe the spin-dependent
transport through the NM-QD-FM junction in the sequential tunneling
regime ($\Gamma_0 \ll k_B T$,\cite{glaz-mat} where $\Gamma_0$ is a
characteristic tunneling rate). An alternative description in terms
of nonequilibrium Green functions is also presented in the appendix,
that corroborates our results obtained via master equation.

\section{Model and Master Equation Approach} 
\begin{figure}[t]
\par
\begin{center}
\epsfig{file=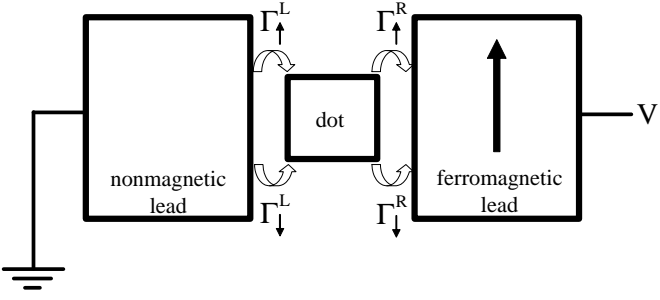, width=0.5\textwidth}
\end{center}
\caption{System studied: a nonmagnetic quantum dot coupled via
tunneling barriers to a nonmagnetic left lead and a ferromagnetic
right lead. A bias voltage $V$ is applied across the system so that
the left ($\mu_L$) and right ($\mu_R$) chemical potentials differ by
$\mu_L-\mu_R=eV$.} \label{fig1}
\end{figure}

The
NM-QD-FM system we study is schematically illustrated in Fig.
\ref{fig1}. An external bias voltage $V$ drives the system away
from equilibrium thus imposing a chemical potential imbalance
between the left (L) and the right (R) leads: $\mu_L-\mu_R=eV$,
where $\mu_{L(R)}$ is the chemical potential of the lead $L(R)$
and $e$ is the absolute value of the electron charge ($e>0$). The
system Hamiltonian is
\begin{eqnarray}
H&=&\sum_{\mathbf{k} \s \eta} \e_{\mathbf{k} \s \eta}
c_{\mathbf{k} \s \eta}^\dagger c_{\mathbf{k} \s \eta}\nonumber\\
&\quad&+\sum_{\s} \e_d d_\s^\dagger d_\s + U d_\uparrow^\dagger d_\uparrow d_\downarrow^\dagger d_\downarrow \nonumber\\
&\quad&+\sum_{\mathbf{k} \s \eta} (t_{\mathbf{k} \s \eta}
c_{\mathbf{k} \eta}^\dagger d_\s +t_{\mathbf{k} \eta}^*
d_\s^\dagger c_{\mathbf{k} \s \eta}),
\end{eqnarray}
where $\e_{\mathbf{k} \s \eta}$ is the free-electron energy with
wave vector $\mathbf{k}$ and spin $\s$ in lead $\eta$ ($\eta=L,
R$), $\e_d$ is the spin-degenerate dot level, $U$ is the on-site
Coulomb repulsion and the operators $c_{\mathbf{k} \s \eta}$
($c_{\mathbf{k} \s \eta}^\dagger$) and $d_\s$ ($d_\s^\dagger$)
destroy (create) an electron with spin $\s$ in the lead $\eta$ and
in the dot, respectively. The matrix element $t_{\mathbf{k} \s
\eta}$ gives the lead-dot coupling. We do not consider any
spin-flip processes.

To calculate the current we use rate equations,\cite{wr01,glaz-mat}
which yield
\begin{eqnarray}\label{Isetageneral}
I_\s^\eta &=&  e [\G_{01
\s}^\eta(1-n_\s-n_{\overline{\s}}+n_{\uparrow \downarrow}) -
\G_{10\s}^\eta (n_\s-n_{\uparrow \downarrow})]\nonumber\\&& +
e[\widetilde{\G}_{01\s}^\eta(n_{\overline{\s}}-n_{\uparrow
\downarrow}) - \widetilde{\G}_{10\s}^\eta n_{\uparrow \downarrow}],
\end{eqnarray}
where we have assumed $\hbar=1$. The parameter $\G_{01\s}^\eta$
corresponds to the rate of adding one electron to the dot coming
from lead $\eta$, and $\G_{10\s}^\eta$ is the rate of moving one
electron from the dot to lead $\eta$. In addition,
$\wid{\G}_{01\s}^\eta$ and $\wid{\G}_{10\s}^\eta$ give the rates of
moving one electron with spin $\s$ to and from the dot,
respectively, when it is already occupied by one electron with
opposite spin. Following Ref. [\onlinecite{wr01}] we define $n_{\s}
= \langle \hat{n}_{\s} \rangle$ and $n_{\uparrow \downarrow}=\langle
\hat{n}_{\uparrow} \hat{n}_{\downarrow} \rangle$
($\hat{n}_\s=d_\s^\dagger d_\s$) as the dot single and double
average occupancies, respectively. The tunneling rates are
\begin{eqnarray}
\G_{01\s}^\eta &=& \G_\s^\eta f_\eta \label{G01s}\\
\G_{10\s}^\eta &=& \G_\s^\eta (1-f_\eta)\label{G10s}\\
\wid{\G}_{01\s}^\eta &=& \wid{\G}_\s^\eta \wid{f}_\eta\label{G01st}\\
\wid{\G}_{10\s}^\eta &=& \wid{\G}_\s^\eta
(1-\wid{f}_\eta),\label{G10st}
\end{eqnarray}
where $f_\eta=1/\{\mathrm{exp}[(\e_d-\mu_\eta)/(k_BT)]+1\}$ and
$\wid{f}_\eta=1/\{\mathrm{exp}[(\e_d+U-\mu_\eta)/(k_BT)]+1\}$. The
rates $\G_\s^\eta$ and $\wid{\G}_\s^\eta$ are related to the
spin-resolved density of states of lead $\eta$ via $\G_\s^\eta=2\pi
|t|^2 \rho_{\s \eta}(\e_d)$ and $\wid{\G}_\s^\eta=2\pi |t|^2
\rho_{\s \eta}(\e_d+U)$. Here we assume
$\Gamma^L_\uparrow=\Gamma^L_\downarrow$ and $\Gamma^R_\uparrow \neq
\Gamma^R_\downarrow$. This reflects the fact that the density of
states of the left lead is spin-degenerate while the right one is
spin-split. Assuming a constant density of states and a constant
tunneling parameter $t$, we have $\G_\s^\eta = \wid{\G}_\s^\eta$.
With this assumption the terms with $n_{\uparrow \downarrow}$ in Eq.
(\ref{Isetageneral}) cancel out,\cite{wr01} and one simply finds
\begin{eqnarray}\label{Iseta}
I_\s^\eta &=&  e [\G_{01\s}^\eta (1-n_\s-n_{\bar{\s}}) -
\G_{10\s}^\eta n_\s + \widetilde{\G}_{01\s}^\eta n_{\bar{\s}}].
\end{eqnarray}

To calculate the current via Eq. (\ref{Iseta}) we need to find
$n_\s$ from\cite{wr01}
\begin{eqnarray}\label{eqns1general}
\frac{d}{dt} n_\s &=& \G_{01\s} [1-n_\s-n_{\bar{\s}}+n_{\uparrow
\downarrow}]-\G_{10\s} [n_\s - n_{\uparrow
\downarrow}]\nonumber\\&&\phantom{xxx}+\wid{\G}_{01 \s}
[n_{\bar{\s}}-n_{\uparrow \downarrow}]-\wid{\G}_{10\s} n_{\uparrow
\downarrow},
\end{eqnarray}
where
\begin{eqnarray}
\G_{01\s}&=& \G_{01\s}^L+\G_{01\s}^R=\G_\s^L f_L + \G_\s^R f_R\\
\G_{10\s}&=& \G_{10\s}^L+\G_{10\s}^R=\G_\s^L(1-f_L)+\G_\s^R(1-f_R)\\
\widetilde{\G}_{01\s}&=&
\widetilde{\G}_{01\s}^L+\widetilde{\G}_{01\s}^R= \wid{\G}_\s^L
\wid{f}_L+\wid{\G}_\s^R \wid{f}_R\\
\widetilde{\G}_{10\s}&=&
\widetilde{\G}_{10\s}^L+\widetilde{\G}_{10\s}^R= \wid{\G}_\s^L
(1-\wid{f}_L)+\wid{\G}_\s^R (1-\wid{f}_R).
\end{eqnarray}
When $\G_\s^\eta=\wid{\G}_\s^\eta$, Eq. (\ref{eqns1general}) becomes
\begin{equation}\label{eqns1}
\frac{d}{dt} n_\s = \G_{01\s} [1-n_\s-n_{\bar{\s}}]-\G_{10\s} n_\s
+\widetilde{\G}_{01 \s} n_{\bar{\s}},
\end{equation}
where the terms with $n_{\uparrow \downarrow}$ cancel out.

\emph{Stationary regime}. In this regime ($dn_\s/dt=0$) Eq.
(\ref{eqns1}) reduces to
\begin{equation}\label{ns_nsbar}
n_\s=\frac{\G_{01\s}+(\widetilde{\G}_{01\s}-\G_{01\s})n_{\bar{\s}}}{\G_{01\s}+\G_{10\s}},
\end{equation}
which can be solved for each spin component, thus resulting in
\begin{equation}\label{nsfinal}
n_\s=\frac{\G_{01\s}\G_{10\bar{\s}}+\G_{01\bar{\s}}\wid{\G}_{01\s}}{\Pi_\s},
\end{equation}
where
$\Pi_\s=(\G_{01\s}+\G_{10\s})(\G_{01{\bar{\s}}}+\G_{10{\bar{\s}}})
-(\wid{\G}_{01\s}-\G_{01\s})(\wid{\G}_{01{\bar{\s}}}-\G_{01{\bar{\s}}})$.
Using Eq. (\ref{nsfinal}) into Eq. (\ref{Iseta}) we obtain
\begin{widetext}
\begin{equation}\label{Ifinal}
I_\s^\eta= e
\frac{\G_{01\s}^\eta(\G_{10\s}\G_{10\bar{\s}}-\wid{\G}_{01\s}\wid{\G}_{01\bar{\s}})
-\G_{10\s}^\eta(\G_{01\s}\G_{10\bar{\s}}+\G_{01\bar{\s}}\wid{\G}_{01\s})
+\wid{\G}_{01\s}^\eta(\G_{01\bar{\s}}\G_{10\s}+\G_{01\s}\wid{\G}_{01\bar{\s}})}
{\Pi_\s}.
\end{equation}
\end{widetext}
From Eq. (\ref{Ifinal}) we can readily evaluate the current
polarization $\wp=(I_\uparrow - I_\downarrow)/(I_\uparrow +
I_\downarrow)$. Next (Sec. III) we provide some simple analytical
results valid when double-occupancy is energetically forbidden.
Numerical results are presented in Sec. IV.

\section{Regime of Singly occupied dot}\label{analytical}

As we shall see, the most interesting behavior takes place when the
channel $\e_d$ is completely within the conduction window and
$\e_d+U$ is far above the Fermi energy of the emitter. With this
channel configuration we approximate $\widetilde{f}_\eta=0$, and
$f_L=1$, $f_R=0$ for $eV>0$ and $f_L=0$, $f_R=1$ for $eV<0$. Using
this into the occupation and current equations, Eqs. (\ref{nsfinal})
and (\ref{Ifinal}), we find analytical expressions for the first
plateau that appears in the current and its polarization for both
positive and negative bias voltage. Equation (\ref{nsfinal}) then
becomes
\begin{equation}\label{nupdwsolved1}
n_\s=\frac{\G_\s^\eta
\G_{\bar{\s}}^{\bar{\eta}}}{\G_{\s}^{\bar{\eta}}
\G_{\bar{\s}}^{\bar{\eta}}+ \G_\s^L \G_{\bar{\s}}^R +
\G_{\bar{\s}}^L \G_{\s}^R},
\end{equation}
where $\eta=L$, $\bar{\eta}=R$ for $eV>0$ and $\eta=R$,
$\bar{\eta}=L$ for $eV<0$. The current of the left lead then
becomes
\begin{equation}\label{closedI}
    I_\s^L = \pm e \frac{\G_{\bar{\s}}^\eta \G_\s^L \G_\s^R}{\G_\s^\eta
    \G_{\bar{\s}}^\eta + \G_\s^L \G_{\bar{\s}}^R + \G_{\bar{\s}}^L \G_\s^R}
\end{equation}
where $\eta=R$ and the $+$ sign corresponds to $eV>0$, while
$\eta=L$ and the $-$ sign to $eV<0$. The right side current is
simply given by $I_\s^R=-I_\s^L$ for a spin-conserving stationary
regime. Equation (\ref{closedI}) gives the (bias-independent)
current in the regime addressed here. For the particular case of
spin-independent tunneling rates, i.e., $\G_\s^L=\G^L$ and
$\G_\s^R=\G^R$, we obtain
\begin{equation}
I^L = I^L_\uparrow + I^L_\downarrow = \left\{
\begin{array}{c}
  2e(\G^L \G^R)/(2\G^L + \G^R),\phantom{xx}eV>0 \\
  -2e(\G^L \G^R)/(\G^L + 2\G^R),\phantom{xx}eV<0, \\
\end{array}
\right.
\end{equation}
in accordance with results already known in the
literature.\cite{glaz-mat,at03yvn96}

Using Eq. (\ref{closedI}) into the definition $\wp=(I_\uparrow -
I_\downarrow)/(I_\uparrow + I_\downarrow)$, we obtain the current
polarization plateau
\begin{equation}\label{somePol}
\wp=\frac{(\G_{\uparrow}^\eta- \G_{\downarrow}^\eta)}
{(\G_{\uparrow}^\eta+ \G_{\downarrow}^\eta)},
\end{equation}
where $\eta=L$ for $eV>0$ and $\eta=R$ for $eV<0$. We model the
tunneling rates by $\G_\uparrow^L=\G_\downarrow^L=\G_0$ and
$\G_{\uparrow(\downarrow)}^R=\G_0(1 \pm p)$, where $p\in[0,1]$ is
the spin polarization degree of the ferromagnetic right
lead\cite{wr01} and $\G_0$ the lead-dot coupling. Within this model
Eq. (\ref{somePol}) gives
\begin{eqnarray}\label{polresults}
\wp=\left\{ \begin{array}{c}
  0,\phantom{xxxx}eV>0 \\
  p,\phantom{xxxx}eV<0. \\
\end{array} \right.
\end{eqnarray}
Thus, when only the level $\e_d$ is within the conduction window,
the current becomes unpolarized for positive bias, while
spin-polarized for negative bias. Therefore the NM-QD-FM junction
functions as a {\it current-polarization diode}.

\section{RESULTS}

\subsection{Parameters}

We assume that the dot level depends on the bias voltage according
to $\e_d=\e_{gate}-x eV$, where $x$ accounts for asymmetric voltage
drops along the left and right tunnel barriers.\cite{wr01,wr03}
$\e_{gate}$ can be controlled via gate voltages. For the numerics we
take $\e_{gate}=0.5$ meV, $\mu_L=0$, $\mu_R=-eV$, $k_B T=212$
$\mu$eV, $U=3$ meV and $\G_0=10 \mu$eV.\cite{ak04,noteR2} In Secs.
B, C and D we assume a symmetric potential drop across the system
with $x=0.5$. In Sec. E we briefly discuss the asymmetric case with
$x \neq 0.5$.

\subsection{Current polarization}

\begin{figure}[tbp]
\par
\begin{center}
\epsfig{file=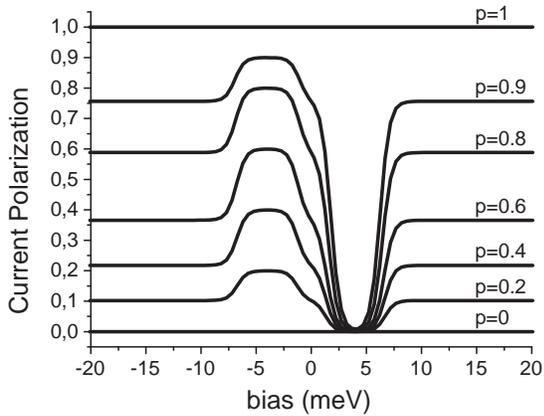, width=0.45\textwidth}
\end{center}
\caption{Current polarization $\wp$ as a function of the external
bias. For $p$ values between $0.2$ and $0.9$, $\wp$ reaches zero for
some particular positive bias range, while for the negative
counterpart it reaches maximum plateaus.} \label{fig2}
\end{figure}

Figure \ref{fig2} shows the current polarization as a function of
the external bias $eV$. We observe that for positive bias the
current polarization decreases for increasing bias, reaching zero
around $eV=$4 meV. Conversely, for the negative biases we obtain a
maximum polarization $p$ around $eV=-4$ meV, confirming the
analytical result found in Sec. III, Eq. (\ref{polresults}). The
voltage range for this behavior scales with the parameter $U$. For
high enough bias voltages ($ |eV| \gtrsim $ 7 meV) the polarization
reaches the same nonzero plateaus for both positive and negative
voltages. Both the suppression ($eV>0$) and the enhancement ($eV<0$)
of the current polarization are due to the interplay of Coulomb
interaction and spin accumulation in the quantum dot. Quite
interestingly this interplay affects $\wp$ differently with the bias
sign, namely, for direct bias it suppresses $\wp$ while for reverse
bias it enhances $\wp$.\cite{fms04these} The suppression of $\wp$
for positive bias results in the zero polarization seen for all $p$
values except $p=1$. In the half-metallic case ($p=1$), there is
only spin up current flowing in the system ($I_\uparrow^\eta \neq
0$, $I_\downarrow^\eta = 0$), so the polarization becomes simply
$\wp=(I_\uparrow^\eta-I_\downarrow^\eta)/(I_\uparrow^\eta+I_\downarrow^\eta)=I_\uparrow^\eta/I_\uparrow^\eta=1$.
On the other hand, for negative bias, the maximum polarization
plateau changes as $p$ varies. In particular, $\wp$ attains a
plateau equal to the polarization degree of the ferromagnetic lead,
according to Eq. (\ref{polresults}). To gain a more detailed
understanding of the spin-diode effect we investigate next the spin
accumulation $m = n_\uparrow - n_\downarrow$ and the spin-resolved
$I-V$ curves as a function of the bias.

\subsection{Spin accumulation}

\begin{figure}[tbp]
\par
\begin{center}
\epsfig{file=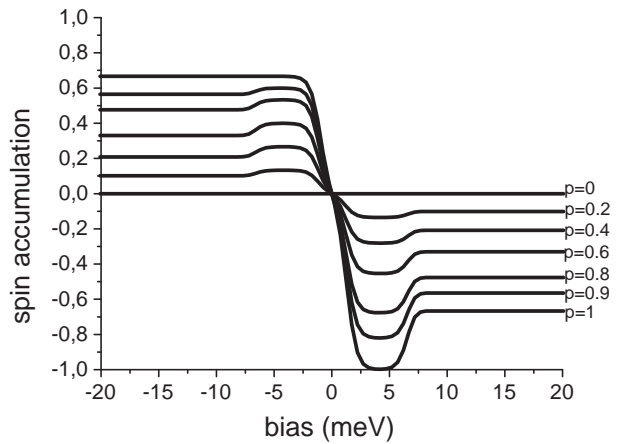, width=0.5\textwidth}
\end{center}
\caption{Spin accumulation $m=n_\uparrow-n_\downarrow$ as a function
of the external bias. For $p=0$ (unpolarized lead) there is no spin
accumulation in the dot. When $p$ increases the spin accumulation
increases as well becoming negative for $eV>0$ and positive for
$eV<0$.} \label{fig3}
\end{figure}

Figure \ref{fig3} shows the spin accumulation
$m=n_\uparrow-n_\downarrow$ as a function of the bias voltage, for
distinct polarization parameters $p$. For all the $p$ values
considered here we note that $m<0$ for positive bias and $m>0$ for
negative bias. This spin-imbalance can be understood in terms of the
tunneling rates $\G_\s^\eta$ between dot and leads. Due to the
ferromagnetism of the right lead, the rates $\G_\s^L$ and $\G_\s^R$
become asymmetric. For example, for $p=0.2$ the rates are
$\G_\uparrow^R=12$ $\mu$eV, $\G_\downarrow^R=8$ $\mu$eV and
$\G_\uparrow^L=\G_\downarrow^L=10$ $\mu$eV. For positive bias,
$\G_\s^L$ becomes the ingoing tunneling rate for electrons with spin
$\s$ and $\G_\s^R$ the outgoing tunneling rate. Due to the
inequality $\G_\uparrow^R > \G_\uparrow^L$, the spin up electrons
can tunnel out the dot faster than they come into it. On the other
hand, since $\G_\downarrow^R < \G_\downarrow^L$, the spin down
electrons leave the dot slower than they come into it. So on average
the spin down electrons spend more time in the dot than the spin up
ones for $eV>0$, thus $n_\downarrow > n_\uparrow \Rightarrow m<0$. A
similar reasoning applies to the other $p$ values, except for $p=0$
for which there is no accumulation. For negative bias,
$\G_\uparrow^L$ and $\G_\downarrow^L$ are the outgoing tunneling
rates while $\G_\uparrow^R$ and $\G_\downarrow^R$ become the ingoing
tunneling rates. As a consequence of this interchange, the spin
accumulation inverts its sign $(m>0)$. For small $p$ values the spin
accumulation is essentially an odd function of the bias, Fig.
\ref{fig3}.

When $p$ increases, though, the imbalance becomes stronger for
positive bias. In particular for $p=1$, $m$ reaches $-1$ in the
positive bias range corresponding to single occupancy
($\e_d+U>\mu_L$), and a constant plateau for all negative bias. This
happens because no spin-down states are available in the right lead
for $p=1$, so a spin-down electron that enters the dot, coming from
the left side ($eV>0$), cannot leave the dot to the right side.
Hence a spin-up electron cannot hop into the dot when
$\e_d+U>\mu_L$, so the accumulation becomes completely spin-down
polarized for positive bias. For high enough bias voltages an
additional electron with opposite spin can jump into the dot (for
both positive and negative bias), thus resulting in a suppression
(in modulus) of $m$.

\subsection{Spin-resolved currents}

\begin{figure}[tbp]
\par
\begin{center}
\epsfig{file=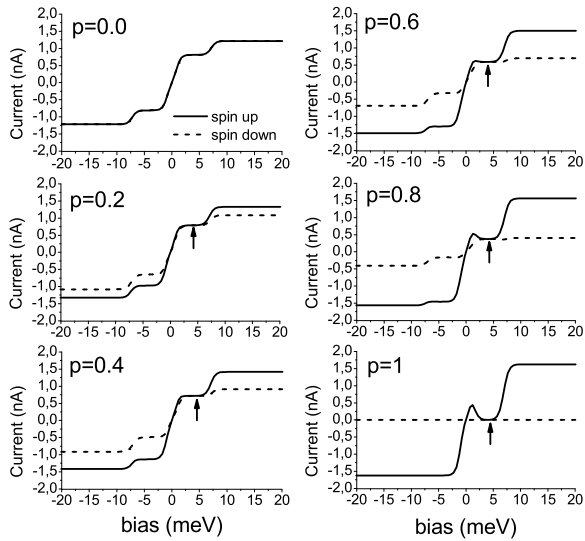, width=0.5\textwidth}
\end{center}
\caption{Spin-resolved currents against bias voltage. For the
plateaus indicated by arrows, $I_\uparrow$ lies on top of
$I_\downarrow$. This gives rise to the $\wp=0$ plateau seen in Fig.
\ref{fig2}. For big enough polarizations (e.g. $p>0.8$) a
negative-differential resistance range arises in the spin up
current.} \label{fig4}
\end{figure}

In Figure \ref{fig4} we show the spin resolved currents $I_\uparrow$
and $I_\downarrow$ as a function of the bias voltage for differing
polarization parameters $p$. We observe that for positive bias the
spin up and spin down currents coincide in the plateaus indicated by
arrows for any $p$ value. This results in the zero
current-polarization seen in Fig. \ref{fig2}. In the second plateau,
though, $I_\uparrow$ attains higher values compared to
$I_\downarrow$, which enhances $\wp$. The strong suppression of
$I_\uparrow$ in the first plateau ($eV>0$) is attributed to the spin
imbalance $m<0$ observed for the corresponding bias range (see Fig.
\ref{fig3}). More specifically, since the dot is predominantly
spin-down occupied for positive bias, the spin-up electrons tend to
be more blocked than the spin-down ones, thus reducing further
$I_\uparrow$ and interestingly locking it on top of $I_\downarrow$.
In contrast, for negative bias we have the population inversion
$m>0$. This gives a stronger suppression of $I_\downarrow$ as
compared to $I_\uparrow$, which enhances the difference between
$I_\uparrow$ and $I_\downarrow$, and consequently $\wp$. When the
channel $\e_d+U$ reaches resonance ($eV \lesssim -7$ meV) both the
$I_\uparrow$ and $I_\downarrow$ plateaus attain values somewhat
closer to each other, thus reducing the current polarization (see
Fig. \ref{fig2}).

In particular for $p=1$ the $I_\downarrow$ is zero for any bias
voltage since there are no spin-down states available in the right
lead. The $I_\uparrow$ increases slightly (for positive bias) while
the dot is becoming populated. When the population is high enough
the Coulomb interaction plays a role and the spin up current goes
down to zero.\cite{refbump} This gives rise to a negative
differential conductance at the beginning of the first plateau for
$eV>0$ [see Fig. \ref{fig4} with $p=1$].\cite{jf05} For negative
bias (and $p=1$) $I_\uparrow$ attains one plateau instead of two
steps as for the other $p$ values. This is expected because the
spin-down electrons do not participate in the transport in this
case, so no Coulomb interaction effect arises. Note that for $p=1$
the system can operate as a mesoscopic current
diode.\cite{iw06,mw05,rs04}

\subsection{Effects of the bias-drop asymmetry}

\begin{figure}[tbp]
\par
\begin{center}
\epsfig{file=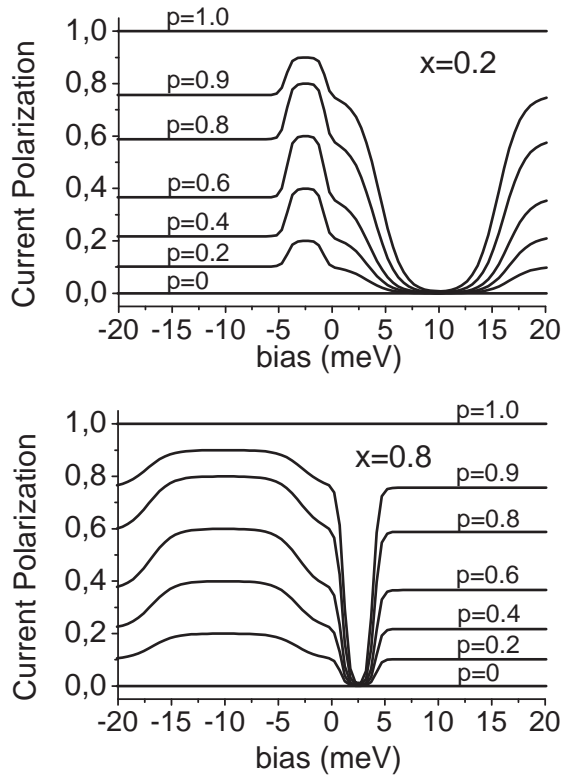, width=0.5\textwidth}
\end{center}
\caption{Current polarization $\wp$ as a function of the bias
voltage for the asymmetry parameters $x=0.2$ (upper panel) and
$x=0.8$ (bottom panel). We observe that the $\wp=0$ plateau enlarges
for $x=0.2$ and shrinks for $x=0.8$ when compared to the $x=0.5$
case. In contrast, the maximum plateau $\wp=p$ reduces for $x=0.2$
and enlarges for $x=0.8$ when compared to the $x=0.5$
case.}\label{fig5}
\end{figure}

Here we consider the effects of an asymmetric bias drop, i.e., $x\ne
0.5$. As Fig. \ref{fig5} shows, the asymmetry in the bias drop gives
rise to quantitative, but not qualitative changes. For $x=0.2$ the
current polarization $\wp$ goes to zero much slower with the bias
than it does for $x=0.5$. This is so because the resonance condition
$\e_d \leq \mu_L$ ($eV>0$), which is necessary to have $\wp=0$ [see
Sec.(\ref{analytical})], happens for higher bias when $x$ decreases.
For negative bias the resonance $\e_d \leq \mu_R$ is reached faster
(i.e., at lower biases as compared with the $x=0.5$ case) for
decreasing $x$. This, in turn, translates into a steeper enhancement
of $\wp$ which then attains a plateau at $\wp=p$ [see Eq.
(\ref{polresults})]. In addition, for $x=0.2$ the zero
current-polarization plateau ($eV>0$) enlarges while the maximum
plateau ($eV<0$) shrinks compared to the respective $x=0.5$ widths.
For $x=0.8$ the resonance $\e_d \leq \mu_L$ ($eV>0$) takes place
faster with the bias when compared to the $x=0.2$ and $x=0.5$ cases.
This results in the steeper suppression of $\wp$ and the shrinkage
of the zero current-polarization bias range. For negative biases the
resonance condition $\e_d \leq \mu_R$ for $x=0.8$ is more slowly
attained with the bias as compared to the $x=0.2$ and $x=0.5$ cases.
Consequently, the polarization $\wp$ reaches the plateau at $\wp=p$
for higher bias voltages (in modulus).

\section{Conclusion}

We propose a NM-QD-FM system which operates as a diode for the
current polarization. More specifically, when double-occupancy is
forbidden in the system, i.e., the channel $\e_d+U$ is far above the
chemical potential of the emitter lead, the system carries an
unpolarized current for positive bias and a spin-polarized current
for negative bias. This effect is a result of the interplay between
spin accumulation in the dot and the Coulomb interaction.
Interestingly, for positive biases the spin-resolved currents
$I_\uparrow$ and $I_\downarrow$ lock onto the same plateau for a
particular bias range, thus resulting in $\wp=0$.

\section{Acknowledgments}

The authors acknowledge G. Platero, K. Flensberg, T. Novotn{\'y},
and J. P. Morten for helpful comments. FMS and JCE acknowledge the
kind hospitality at the Centre for Advanced Study (Oslo) during the
revision stage of this work. JCE acknowledges support from CNPq and
FAPESP.

\appendix

\section{Nonequilibrium Green functions}

In the main text we have formulated the problem via master equation
formalism. Here we show that Eqs. (\ref{nupdwsolved1}) and
(\ref{closedI}), from which our main result Eq. (\ref{polresults})
directly follows, can be obtained via the Keldysh formalism. We
start with the well known equation for the stationary
current\cite{MW92,apj94}
\begin{equation}\label{currentGF}
    I_\s^\eta=ie \int \frac{d\e}{2\pi}
    \{\Gamma_\s^\eta\{[G_{\s\s}^r(\e)-G_{\s\s}^a(\e)]f_\eta(\e)+G_{\s\s}^<(\e)\}\}\nonumber,
\end{equation}
where $G^r_{\s\s}$, $G^a_{\s\s}$ and $G^<_{\s\s}$ are the retarded,
advanced and lesser Green functions, respectively. To calculate
these we apply the equation of motion technique and use the
Hartree-Fock approximation to factorize high-order correlation
functions in the resulting chain of equations.\cite{hh96} The
retarded Green function becomes
\begin{equation}\label{Gdot}
G_{\s\s}^r(\e)=\frac{1}{g_{\s\s}^{-1}(\e)-\Sigma_{\s}^{r}(\e)},
\end{equation}
where $\Sigma_\s^r(\e)$ is the non-interacting tunneling self-energy
given in the wide band approximation by $\Sigma_\s^r(\e)=-i \G_\s /
2=-i(\G_\s^L+\G_\s^R)/2$, and $g_{\s\s}(\e)$ is the dot Green
function without coupling to leads,
\begin{equation}\label{smallGdot}
g_{\s\s}(\e)=\frac{\e-\e_d-U(1-n_{\bar{\s}})}{(\e-\e_d)(\e-\e_d-U)},
\end{equation}
where $n_{\bar{\s}}$ is the dot occupation number, with
$\bar{\s}=-\s$. This occupation can be calculated self-consistently
via
\begin{equation}\label{nintGlesser}
n_\s=\langle d_\s^\dagger d_\s \rangle=-i\int
\frac{d\e}{2\pi}G_{\s\s}^<(\e),
\end{equation}
where the correlation function $G_{\s\s}^<(\e)$ is given by the
Keldysh equation
\begin{equation}
G_{\s\s}^<(\e)=G_{\s\s}^r(\e)\S_\s^<G_{\s\s}^a(\e).
\end{equation}
The advanced Green function $G_{\s\s}^a(\e)$ is given by
$G_{\s\s}^a(\e)=[G_{\s\s}^r(\e)]^*$, while $\S_\s^<=i[\G_\s^L f_L +
\G_\s^R f_R]$. In order to consider the same channel configuration
adopted in Sec. III, we assume a large $U$ ($\e_d+U \gg \mu_\eta$)
so the retarded Green function becomes
\begin{equation}\label{GdotlargeU}
G_{\s\s}^r(\e)=\frac{(1-n_{\bar{\s}})}{\e-\e_d-\Sigma_{\s}^{r}(1-n_{\bar{\s}})},
\end{equation}
and the lesser Green function reads
\begin{equation}\label{GlesserlargeU}
G^<_{\s\s}(\e)=\frac{i(\G_\s^L f_L + \G_\s^R
f_R)(1-n_{\bar{\s}})^2}{(\e-\e_d)^2+(\frac{\G_\s}{2})^2(1-n_{\bar{\s}})^2}.
\end{equation}
Substituting Eq. (\ref{GlesserlargeU}) into Eq. (\ref{nintGlesser})
we have
\begin{equation}\label{nsintegral}
n_\s=\frac{(1-n_{\bar{\s}})^2}{2\pi}\int d\e \frac{(\G_\s^L f_L +
\G_\s^R f_R)}{(\e-\e_d)^2+(\frac{\G_\s}{2})^2(1-n_{\bar{\s}})^2}.
\end{equation}
Now assuming that the dot level $\e_d$ is completely on resonance
within the conduction window between $\mu_L$ and $\mu_R$ for
positive or negative bias\cite{approximation} we can integrate Eq.
(\ref{nsintegral}) in order to obtain
\begin{equation}\label{nsafterint}
n_\s=\frac{\G_\s^\eta}{\G_\s}(1-n_{\bar{\s}}),
\end{equation}
where $\eta=L$ for $eV>0$ or $\eta=R$ for $eV<0$. Solving Eq.
(\ref{nsafterint}) for each spin component we find exactly Eq.
(\ref{nupdwsolved1}).

To obtain Eq. (\ref{closedI}) from the Green functions, we
substitute Eqs. (\ref{GdotlargeU}) and (\ref{GlesserlargeU}) into
the current formula (\ref{currentGF}), which gives
\begin{eqnarray}\label{closedINEGF}
    I_\s^L&=& e\int \frac{d\e}{2\pi} \frac{ \G_\s^L \G_\s^R
    (1-n_{\bar{\s}})^2(f_L-f_R)}{(\e-\e_d)^2+(\frac{\G_\s}{2})^2(1-n_{\bar{\s}})^2}.
\end{eqnarray}
Solving Eq. (\ref{closedINEGF}) with the same assumptions adopted
previously\cite{approximation} we find
\begin{equation}\label{IsLafterint}
I_\s^L = \pm e \frac{\G_\s^L \G_\s^R}{\G_\s} (1-n_{\bar{\s}}),
\end{equation}
where $+$ and $-$ signs correspond to $eV>0$ and $eV<0$,
respectively. Eq. (\ref{IsLafterint}) can also be written as Eq.
(\ref{closedI}).


\begin{thebibliography}{99}
\bibitem{noteR2} In the context of nonequilibrium Green function (see Appendix) we note
that by increasing the coupling parameter $\G_0$, the width of the
dot level increases as well, thus resulting in more smooth curves in
Figs. \ref{fig2}-\ref{fig5}.
\bibitem{approximation} More precisely we assume that $\mu_R<\e_d<\mu_L$
for $eV>0$ and $\mu_L<\e_d<\mu_R$ for $eV<0$. In addition we
approximate $f_L \approx 1$ and $f_R \approx 0$ for $eV>0$ and
$f_L \approx 0$ and $f_R \approx 1$ for $eV<0$.
\end{thebibliography}
\end{document}